\documentclass[aps,preprintnumbers,nofootinbibt,referee, keywords,referee]{revtex4}
\usepackage{graphicx}
\usepackage{amsmath}
\usepackage{amssymb}
\usepackage{amsfonts}
\usepackage{bm}
\usepackage{color}

\def\be{\begin{equation}}
\def\ee{\end{equation}}
\def\bea{\begin{eqnarray}}
\def\eea{\end{eqnarray}}

\begin{document}

\title{New integrability case for the Riccati equation}
\author{M. K. Mak}
\email{mkmak@vtc.edu.hk}
\affiliation{Department of Physics and Center for Theoretical and Computational Physics,
The University of Hong Kong, Pok Fu Lam Road, Hong Kong, P. R. China}
\author{T. Harko}
\email{harko@hkucc.hku.hk}
\affiliation{Department of Physics and Center for Theoretical and Computational Physics,
The University of Hong Kong, Pok Fu Lam Road, Hong Kong, P. R. China}

\begin{abstract}
A new integrability condition of the Riccati equation $dy/dx=a(x)+b(x)y+c(x)y^{2}$ is presented. By introducing an auxiliary equation depending on a generating function $f(x)$, the general solution of the Riccati equation can be obtained if the coefficients $a(x)$, $b(x)$, $c(x)$, and the function $f(x)$ satisfy a particular constraint.  The validity and reliability of the method are tested by obtaining the  general solutions of some  Riccati type differential equations. Some applications of the integrability conditions for the case of the damped harmonic oscillator with time dependent frequency, and for solitonic wave, are briefly discussed.

\end{abstract}

\maketitle

\section{Introduction}

The Riccati equations, of  the type
\begin{equation}
\frac{dy}{dx}=a(x)+b(x)y+c(x)y^{2},  \label{1}
\end{equation}%
where $a$, $b$, $c$ are arbitrary real functions of $x$, with $a,b,c\in C^{\infty}(I)$, defined on a real interval $I\subseteq\Re$  \cite{1r,2r},
find surprisingly many applications in physics and mathematics.
For example, supersymmetric quantum mechanics \cite{1}, variational calculus \cite{2}, nonlinear physics \cite{3}, renormalization group equations for running coupling constants in
quantum field theories \cite{4,5}, thermodynamics \cite{6}, the formulation of Newton's law \cite{7}, the dynamical rate equations for physical processes driven by a combination of diffusive growth \cite{8}, obtaining analytical solutions to the (3+1)-dimensional Gross-Pitaevskii equation in the presence of chirp and for different
diffraction and potential functions \cite{9}, and the study of the cubic nonlinear Ginzburg-Landau equations \cite{8a}   are just a
few topics where Riccati equations play a key role. One of the main
reason for their importance in physical applications is that a change of function turns the Riccati equation into a linear second-order differential equation, that stands as basic mathematical background for many areas
of physics.  Group theoretical methods, which are  very useful for a better understanding of the properties of the Riccati equation, and a discussion of the integrability conditions from a group theoretical perspective can be found in \cite{Ram}. From a group theoretical point of view the nonlinear superposition principle also arises in a simple way.

Since the Riccati equation is a widely studied nonlinear equation, knowing that the physical system under consideration can be brought into Riccati form has certainly many advantages in the investigation of its properties. If one or two particular solutions $y^p_{i}(x)$, $i=1,2$ of the Riccati equation are known, the equation may be solved by quadratures \cite{1r,2r}.

A number of solutions of the Riccati equation can be obtained by assuming that the coefficients $a(x)$, $b(x)$, and $c(x)$ satisfy some particular constraints. Thus, if the coefficients $a$, $b$, $c$, defined  and   continuous  in  some  interval $I\subset \Re$, are related as
\be
a+b+c=\frac{d}{dx}\log\frac{\alpha }{\beta }-\frac{\alpha -\beta}{\alpha \beta }\left(\alpha c-\beta a\right),
\ee
 with $\alpha (x)$  and  $\beta (x)$  properly chosen functions differentiable in $I$,  such that  $\alpha \beta > 0$, then the Riccati equation is integrable  by  quadratures \cite{Stel}. If $c(x)\equiv 1$ and the functions $a(x)$ and $b(x)$ are polynomials satisfying the condition
 \be\label{integ}
 \Delta =b^2(x)-2\frac{db(x)}{dx}-4a(x)\equiv {\rm constant},
 \ee
 then
 \be
 y_1(x)=-\frac{\left[b(x)+\sqrt{\Delta}\right]}{2},
 \ee
and
\be
y_2(x)=-\frac{\left[b(x)-\sqrt{\Delta}\right]}{2},
\ee
are both solutions of the Riccati equation \cite{1r,2r}. Note that the choices $c(x)\equiv 1$ and $b(x)$ and $c(x)$ being of polynomial form are restrictive conditions of the method \cite{1r,2r}. If we know three particular solutions $y^p_i(x)$, $i=1,2,3$, then the Riccati equation can be solved without quadratures \cite{1r,2r}.

It is the purpose of the present paper to introduce a new integrability case for the Riccati equation. By introducing an auxiliary equation depending on a generating function $f(x)$, the general solution of the Riccati equation can be obtained if the coefficients $a(x)$, $b(x)$, $c(x)$, and the function $f(x)$ satisfy a particular constraint.  The validity and reliability of the method are tested by obtaining the  general solutions of some  Riccati type differential equations. Some applications of the integrability conditions for the case of the damped harmonic oscillator with time dependent frequency, and for solitonic wave, are briefly discussed.

The present paper is organized as follows. The integrability condition for the Riccati equation is obtained in Section~\ref{sect2}. Some specific examples of the integrability condition obtained by fixing the functions $b(x)$, $c(x)$ and $f(x)$ are shown in Section~\ref{sect3}. Some integrable Riccati equations with fixed $a(x)$, $b(x)$ and $f(x)$ are presented in Section~\ref{sect4}. The physical applications of the method are briefly outlined in Section~\ref{sect5}. We conclude our results in Section~\ref{sect6}.

\section{The integrability condition for the Riccati equation}\label{sect2}

From an algebraic point of view Eq.~(\ref{1}) is a quadratic equation in $y$. We consider that its particular solutions $y_{\pm}^{p}(x)$ can be represented as
\begin{equation}\label{y0}
y_{\pm}^{p}(x)=\frac{-b(x)\pm \sqrt{f(x)}}{2c\left( x\right) },
\end{equation}
where we have introduced the new function $f(x)\in C^{\infty }(I)$, defined as
\begin{equation}
f\left( x\right) =b^{2}\left( x\right) -4c\left( x\right) \left[ a\left(
x\right) -\frac{dy}{dx}\right].
\end{equation}

The requirement that $y_{\pm }^{p}(x)$ as defined in Eq.~(\ref{y0}) is a solution of the Riccati Eq.~(\ref{1}), restricts the expression of $a(x)$ to the form
\begin{equation}\label{4_1}
a(x)=\frac{d}{dx}\left[ \frac{-b(x)\pm \sqrt{f(x)}}{2c(x)}\right] +\frac{%
b^{2}(x)-f(x)}{4c(x)}.
\end{equation}

By substituting $a(x)$ given by Eq.~(\ref{4_1}) into Eq.~(\ref{1}), we obtain an auxiliary Riccati
equation of the form
\begin{eqnarray}
\frac{dy}{dx}=\frac{d}{dx}\left[ \frac{-b(x)\pm \sqrt{f(x)}}{2c(x)}\right]
+\frac{b^{2}(x)-f(x)}{4c(x)}+
b(x)y+c(x)y^{2},  \label{2}
\end{eqnarray}
where $f(x)$ is a solution generating function to the auxiliary Riccati
Eq.~(\ref{2}). Therefore we obtain the following

{\bf Theorem} The general solution of Eq.~(\ref{2}) with particular solutions given by Eq.~(\ref{y0}) is represented by
\begin{eqnarray}\label{6_1}
y_{\pm}(x)=e^{\pm \int \sqrt{f(x)}dx}\left[ -\int c(x)e^{\pm \int \sqrt{f(x)}%
dx}dx+C_{\pm}\right] ^{-1}+
\left[ \frac{-b(x)\pm \sqrt{f(x)}}{2c(x)}\right] ,
\end{eqnarray}
where $C_{\pm}$ is an arbitrary integration constant.

Hence by fixing the functional forms of the functions $b(x)$, $c(x)$ and $f(x)$, we can obtain the general solution of the Riccati Eq.~(\ref{2}) with $a(x)$ given by Eq.~(\ref{4_1}). Alternatively, by fixing the functional forms of the functions $a(x)$, $b(x)$ and $f(x)$, we can obtain the general solution of the Riccati Eq.~(\ref{2}), with $c(x)$ given by Eq.~(\ref{4_1}).

In the case $c(x)\equiv 1$ and $f(x)\equiv \Delta ={\rm constant}$, from Eq.~(\ref{4_1}) we reobtain the integrability condition given by Eq.~(\ref{integ}).

\section{Generating solutions of the Riccati equation by fixing $b(x)$, $c(x)$ and $f(x)$}\label{sect3}

\subsection{Solutions with $f(x)\equiv 0$}

In the case $f(x)\equiv 0$ Eq.~(\ref{4_1}) becomes
\begin{equation}\label{4}
a(x)=\frac{d}{dx}\left[ \frac{-b(x)}{2c(x)}\right] +\frac{%
b^{2}(x)}{4c(x)},
\end{equation}
and from Eq.~(\ref{6_1}) it follows that the Riccati  Eq.~(\ref{2}) satisfying the condition given by Eq.~(\ref{4}) has the general solution
\be\label{7a}
y(x)=\left[ -\int c(x)dx+C\right] ^{-1}+
\left[ \frac{-b(x)}{2c(x)}\right].
\ee

Thus by giving the functional form of the two functions $b(x)$ and $c(x)$ the general solutions of the Riccati equation can be obtained.

\subsubsection{ Example 1.}

The coefficients in the Riccati equation
\be\label{ex1}
\frac{dy}{dx}=
\frac{1}{4} e^{(\beta -\alpha )x } x^{m-n-1}
\left[-2 m+2 n+x
   \left(e^{\beta x } x^m+2 \alpha -2 \beta \right)\right]+e^{\beta x}x^my+e^{\alpha x}x^ny^2,
\ee
where $\alpha $, $\beta $, $n$, $m$ are arbitrary real constants,  satisfy Eq.~(\ref{4}). The equation has the particular solution
\be
y^p(x)=-\frac{1}{2} e^{\left(\beta - \alpha \right)x } x^{m-n},
\ee
and the general solution of Eq.~(\ref{ex1}), which follows from Eq.~(\ref{7a}), is given by
\be
y(x)=\frac{1}{x^{n+1}\left(-\alpha x\right)^{-1-n}\Gamma (1+n,-\alpha x ) +C}-\frac{1}{2} e^{\left(\beta -
   \alpha \right)x} x^{m-n},
\ee
where $\Gamma (a,z)=\int_z^{\infty }{t^{a-1}e^{-t}dt}$ is the incomplete gamma function \cite{Abram}.

\subsubsection{ Example 2.}

The Riccati equation with Bessel $J_n(x)$ function coefficients,  given by
\begin{eqnarray}\label{12}
\frac{dy}{dx} &=&\frac{x^{-\alpha +\beta -1} \left\{2 x J_{m-1}(x) J_n(x)+J_m(x) \left[2
   (-m+n+\alpha -\beta ) J_n(x)+x \left(x^{\beta } J_n(x){}^2-2
   J_{n-1}(x)\right)\right]\right\}}{4 J_m(x){}^2}+\nonumber\\
   &&x^{\beta }J_{n}(x)y+x^{\alpha }J_{m}(x)y^{2},
\end{eqnarray}
where $\alpha$, $\beta $, $m$, and $n$, are arbitrary real constants, has the particular solution
\be
y^p(x)=-\frac{x^{\beta -\alpha } J_n(x)}{2 J_m(x)}.
\ee

The general solution of Eq.~(\ref{12}) is given by
\begin{widetext}
\begin{equation}
y(x) =-\frac{x^{\beta -\alpha }J_{n}(x)}{2J_{m}(x)}+
\frac{1}{C-2^{-m-1}x^{m+\alpha +1}\Gamma \left[ \frac{1}{2}(m+\alpha
+1)\right] \,_{1}\tilde{F}_{2}\left[ \frac{1}{2}(m+\alpha +1);m+1,\frac{1}{2}%
(m+\alpha +3);-\frac{x^{2}}{4}\right] },
\end{equation}
\end{widetext}
where $\Gamma (z)=\int_0^{\infty}{t^{z-1}e^{-t}dt}$ is the Euler gamma function, and ${}_p\tilde{F}_{q}\left(a;b;z\right)={}_pF_{q}\left(a;b;z\right)/\left[\Gamma \left(b_1\right)....\Gamma \left(b_q\right)\right]$ is the regularized generalized hypergeometric function \cite{Abram}.

\subsection{Solutions with $f(x)\equiv $ constant}

By choosing $f(x)\equiv K^2\equiv $ constant, with $K\in \Re$, Eq.~(\ref{4_1}) becomes
\be\label{15}
a(x)=\frac{d}{dx}\left[ \frac{-b(x)\pm K}{2c(x)}\right] +\frac{%
b^{2}(x)-K^2}{4c(x)},
\end{equation}
and from Eq.~(\ref{6_1}) it follows that the general solution of the Riccati Eq.~(\ref{2}) satisfying the condition given by Eq.~(\ref{15}) is represented by
\be
y_{\pm}(x)=e^{\pm Kx}\left[ -\int c(x)e^{\pm Kx}dx+C_{\pm}\right] ^{-1}+
\left[ \frac{-b(x)\pm K}{2c(x)}\right].
\ee

\subsubsection{ Example 3.}

The Riccati equation
\bea\label{17}
\frac{dy}{dx} &=&\frac{1}{4}%
e^{-\alpha x}x^{-n-1}
\Bigg\{ e^{\beta x}x^{m}\left[ -2m+2n+x\left( e^{\beta x}x^{m}+2\alpha -2\beta
\right) \right] -
K^{2}x\mp 2K(n+\alpha x)\Bigg\}+\nonumber\\
&&e^{\beta x}x^{m}y+ e^{\alpha x}x^{n}y^{2},
\eea
has the particular solution
\begin{equation}
y_{\pm}^{p}(x)=\frac{1}{2}e^{-\alpha x}x^{-n}\left( \pm K-e^{\beta x}x^{m}\right) .
\end{equation}

The general solution of Eq.~(\ref{17}) is given by
\begin{eqnarray}
y_{\pm }(x) =\frac{1}{2} x^{-n}e^{-\alpha x}\left( \pm K-e^{\beta x}x^{m}\right)+
\frac{e^{\pm Kx}}{x^{n+1}\left[-x(\pm K+\alpha
)\right]^{-n-1}\Gamma \left[n+1,-x(\pm K+\alpha )\right]+C_{\pm}}.\nonumber\\
\end{eqnarray}

\subsubsection{ Example 4.}

The Riccati  equation
\begin{eqnarray}\label{20}
\frac{dy}{dx} &=&\frac{1}{4}%
e^{-\alpha x}\csc (nx)
\Bigg\{ e^{\beta x}\Bigg[ \sin (mx)\Bigg( 2\alpha -2\beta +2n\cot
(nx)+
e^{\beta x}\sin (mx)\Bigg) -
2m\cos (mx)\Bigg] \mp\nonumber\\
&& K\left[ \pm K+2\alpha
+2n\cot (nx)\right] \Bigg\}+e^{\beta x}\sin (mx)y+e^{\alpha x}\sin (nx)y^{2},
\end{eqnarray}
has the particular solution
\be
y^{p}_{\pm}(x)=\frac{1}{2} e^{- \alpha x} \csc (n x) \left[\pm K-e^{\beta x } \sin (m
   x)\right].
\ee
The general solution of Eq.~(\ref{20}) is given by
\bea
y_{\pm }(x)=\frac{1}{2} e^{- \alpha x } \csc (n x) \left[\pm K-e^{\beta x } \sin (m
   x)\right]+
\frac{e^{\pm K x}}{C_{\pm}-\frac{e^{(\pm K+\alpha )x }
   \left[(\pm K+\alpha ) \sin (n x)-n \cos (n x)\right]}{K^2\pm 2 \alpha  K+n^2+\alpha
   ^2}}.
\eea

\subsection{Solutions with arbitrary function  $f(x)$}

We choose $f(x)=b^{2}(x)$, that
is, $b(x)=\pm \sqrt{f(x)}$. In this case the Riccati Eq.~(\ref{2}) becomes
\begin{equation}\label{ber}
\frac{dy}{dx}=b(x)y+c(x)y^{2},
\end{equation}
with the general solution obtained with the use of Eq.~(\ref{6_1}) given by
\begin{equation}
y(x)=e^{\int b(x)dx}\left[ -\int c(x)e^{\int b(x)dx}+C\right] ^{-1}.
\end{equation}

The same result can be obtained by directly solving the Bernoulli type  Eq.~(\ref{ber}).

\subsubsection{ Example 5.}

The Riccati  equation
\begin{eqnarray}\label{23}
\frac{dy}{dx} =
\frac{1}{4}x^{-n-1}\left[ -2(m-n)x^{m}+x^{2m+1}-x^{2}+(1-2n)\sqrt{x}\right]+x^{m}y+x^{n}y^{2},
\end{eqnarray}
has the particular solution
\begin{equation}
y^{p}(x)=\frac{1}{2}x^{-n}\left( \sqrt{x}-x^{m}\right) ,
\end{equation}
corresponding to $f(x)=x$. The general solution of Eq.~(\ref{23}) is given by
\bea
y(x)=\frac{1}{2}\left( \sqrt{x}-x^{m}\right) x^{-n}+
\frac{e^{\frac{2x^{3/2}}{%
3}}}{\left( \frac{2}{3}\right) ^{(1-2n)/3}x^{n+1}\left( -x^{3/2}\right) ^{-\frac{2%
}{3}(n+1)}\Gamma \left[
\frac{2(n+1)}{3},-\frac{2x^{3/2}}{3}\right] +C}.
\eea

\subsubsection{ Example 6.}
The Riccati equation,
\be\label{26}
\frac{dy}{dx} =
\frac{1}{4}x^{-n-1}\Bigg[ -\left( x^{n+1}+2m-2n\right)
x^{m}+x^{2m+1}+
(m-n)x^{(m+n)/2}\Bigg] +x^{m}y+x^{n}y^{2},
\ee
 has the particular solution
\begin{equation}
y^{p}(x)=\frac{1}{2}x^{-n}\left[ x^{(m+n)/2}-x^{m}\right] ,
\end{equation}
corresponding to $f(x)=b(x)c(x)=x^{m+n}$. The general solution of Eq.~(\ref{26}) is given by
\begin{equation}
y(x)=\frac{1}{2}\left[  x^{(m+n)/2}-x^{m}\right] x^{-n}+\frac{e^{ \frac{%
2x^{1+(m+n)/2}}{m+n+2}}}{C-\int e^{\frac{2x^{1+(m+n)/2}}{m+n+2}%
}x^{n}\,dx}.
\end{equation}

\section{Generating solutions of the Riccati equation by fixing $a(x)$, $b(x)$ and $f(x)$}\label{sect4}

Integrating Eq. (\ref{4_1}) we obtain
\begin{equation}\label{30}
c_{\pm}(x)=\frac{1}{2}I_{\pm }(x)\left[ b(x)\mp \sqrt{f(x)}\right] \left[ -\int
a(x)I_{\pm }(x)dx+k_{\pm}\right] ^{-1},
\end{equation}%
where $k_{\pm}$ is an arbitrary integration constant, and
\begin{equation}
I_{\pm }(x)=\exp \left\{ -\frac{1}{2}\int {\left[ b(x)\pm \sqrt{f(x)}\right]
dx}\right\} .
\end{equation}

With this form of $c(x)$ the Riccati  Eq.~(\ref{1}) can be written as
\bea\label{Ricc1}
\frac{dy}{dx}=a(x)+b(x)y+\frac{1}{2}I_{\pm
}(x)\left[ b(x)\mp \sqrt{f(x)}\right]
\left[ -\int a(x)I_{\pm }(x)dx+k_{\pm}\right] ^{-1}y^{2}.
\eea

Then the general solution to   Eq.~(\ref{Ricc1}) is
\begin{widetext}
\begin{eqnarray}
y_{\pm}(x) &=&\frac{e^{\pm \int \sqrt{f(x)}dx}}{ -\frac{1}{2}\int I_{\pm }(x)\left[ b(x)\mp
\sqrt{f(x)}\right]
\left[ -\int a(x)I_{\pm }(x)dx+k_{\pm}\right] ^{-1}e^{\pm \int \sqrt{f(x)}dx}dx+C_{\pm}%
} -  \nonumber \\
&&I_{\pm }^{-1}(x)\left[ -\int a(x)I_{\pm }(x)dx+k_{\pm}\right] .
\end{eqnarray}
\end{widetext}

\subsubsection{Example 7.}

The Riccati equation
\be\label{35}
\frac{dy}{dx}=\frac{\alpha }{x^m}+\frac{\beta}{x^m} y+\frac{\beta ^2 x^{-m} }{4 \alpha +2 k \beta e^{\frac{\beta x^{1-m} }{2(1- m)}}  }y^2,
\ee
where $k\in\Re$ is a constant, whose coefficients satisfy the condition given by Eq.~(\ref{30}) with $f(x)\equiv 0$, has the particular solution
\be
y^p(x)=-\frac{2 \alpha }{\beta }-ke^{\frac{\beta x^{1-m} }{2(1- m)}} .
\ee

The general solution of Eq.~(\ref{35}) is given by
\be
y(x)=-\frac{2 \alpha }{\beta }-ke^{\frac{\beta x^{1-m}  }{2(1- m)}}
   +\frac{4 (m-1) \alpha  x^m}{2 (m-1) \left[2 C \alpha +\beta
   \log \left(2 \alpha +k \beta e^{\frac{\beta x^{1-m}  }{2(1- m)}}
   \right)\right] x^m+\beta ^2 x}.
\ee

\subsubsection{Example 8.}

The Riccati equation
\be\label{40}
\frac{dy}{dx}=x^{m/2}e^{x/2}+\frac{m}{x}y+\frac{1}{2}x^{-1-m/2}e^{-x/2}y^2,
\ee
whose coefficients satisfy the condition given by Eq.~(\ref{30}) with $f(x)\equiv 1$, $a(x)=I_{\pm}^{-1}(x)$, and $k=m$, respectively, has the particular solution
\be
y^p(x)=-mx^{m/2}e^{x/2}.
\ee
The general solution of Eq.~(\ref{40}) is given by
\be
y(x)=x^{m/2}e^{x/2} (x-m) +\frac{e^x}{\frac{1}{2}
   x^{-m/2}E_{m/2+1}\left(-x/2\right) +C},
\ee
where $E_n(z)$ is the exponential integral function $E_n(z)=\int_1^{\infty}{e^{-zt}/t^ndt}$.

\subsubsection{Example 9.}

The Riccati equation
\be\label{50}
\frac{dy}{dx}=\frac{\beta }{x}+\frac{\alpha }{x}y+\frac{\alpha ^2-1}{2 \left[ 2
   \beta  x+(\alpha +1)k x^{\frac{\alpha +3}{2}}\right]}y^2,
\ee
whose coefficients satisfy the condition given by Eq.~(\ref{30}) with $f(x)=x^{-2}$, has the particular solution
\be
 y^p(x)=\frac{\alpha  \left[ (\alpha +1)k x^{\frac{\alpha +1}{2}}+2 \beta
   \right]}{1-\alpha ^2}.
 \ee
 The general solution of Eq.~(\ref{50}) is given by
 \be
 y(x)=-\frac{2 \beta }{\alpha +1}-k x^{\frac{\alpha +1}{2}}-\frac{4 \beta  x}{\left(\alpha
   ^2-1\right)x  \, _2F_1\left[\frac{2}{\alpha +1},1;\frac{\alpha
   +3}{\alpha +1};-\frac{(\alpha +1)k x^{\frac{\alpha +1}{2}} }{2
   \beta }\right]-4 C \beta },
   \ee
where $\, _2F_1(a,b;c;z)$ is the hypergeometric function.

\section{Applications in physics}\label{sect5}

In the following we consider two physical application of the new
integrability conditions for the Riccati equation

\subsection{The damped time-dependent  harmonic oscillator   }

Many natural  processes can be modeled in the classical regime by the motion
of a damped particle in a time-dependent harmonic  potential, described by
the equation \cite{osc}
\begin{equation}
\ddot{x}+\gamma (t)\dot{x}+\omega ^{2}(t)x=0,  \label{ric}
\end{equation}
where a dot denotes the derivative with respect to the time $t$, $x(t)$ is the
position of the particle, $\gamma (t)$ is the damping function, and $\omega
^{2}(t)$ is the time dependent angular frequency, respectively.  Riccati parameter families of damping modes, related to the Newtonian free damping ones by means of Witten¡¦s supersymmetric scheme were considered in \cite{Rosu1}. This procedure leads to one parameter families of transient modes for each of the three types of free damping, corresponding to a particular type of anti-restoring acceleration. The Ermakov-Lewis procedure was applied to the one-parameter damped modes introduced in \cite{Rosu1} in \cite{Rosu2}.

By introducing the transformation $u=\dot{x}/x$, $x(t)=x_{0}\exp \left[
\int_{0}^{t}u\left( t^{\prime }\right) dt^{\prime }\right] $, where $%
x_{0}=x\left( 0\right) $,  Eq.~(\ref{ric}) can be transformed into a Riccati
type equation, given by
\begin{equation}
\dot{u}=-\omega ^{2}(t)-\gamma (t)u-u^{2}.  \label{ric1}
\end{equation}

Eq.~(\ref{ric1}) has three general solutions, corresponding to three
different constraints imposed on the damping and angular frequency
functions. If the two functions $\omega (t)$ and $\gamma (t)$ satisfy the condition
\begin{equation}
\omega ^{2}(t)=\frac{\dot{\gamma}(t)}{2}+\frac{\gamma ^{2}(t)}{4},
\end{equation}
the general solution of  Eq.~(\ref{ric1}) is given by
\begin{equation}
u(t)=-\frac{\gamma (t)}{2}+\frac{1}{C+t}.
\end{equation}

The integration constant $C$ can be determined from the initial condtion $%
u(0)=\dot{x}(0)/x(0)=v_{0}/x_{0}$, where $v_{0}$ is the initial velocity of
the particle. Therefore the integration constant is
\begin{equation}
C=\frac{2x_{0}}{2v_{0}+\gamma _{0}x_{0}},
\end{equation}
where $\gamma _{0}=\gamma (0)$. The general solution of Eq.~(\ref{ric})  is
given by
\begin{eqnarray}
x(t)=x_0\left(1+\frac{t}{C}\right)\exp\left[-\frac{1}{2}\int_0^t{\gamma \left(t'\right)dt'}\right].
\end{eqnarray}

If the condition
\begin{equation}
\omega ^{2}(t)=\frac{\dot{\gamma}(t)}{2}+\frac{\gamma ^{2}(t)-K^2}{4},
\end{equation}
is satisfied, where $K$ is a constant, the general solution of Eq.~(\ref
{ric1}) is given by
\begin{equation}
u_{\pm}(t)=-\left[\frac{\gamma (t)\pm K}{2}\right]+\frac{\exp\left(\pm Kt\right)}{C_{\pm }\pm \left(1/K\right)\exp(\pm Kt)},
\end{equation}
and the general solution of Eq.~(\ref{ric}) is given by
\begin{equation}
x_{\pm }\left( t\right) =\frac{x_{0}}{\left( C_{\pm }\pm K^{-1}\right) }%
e^{-(1/2)\int_{0}^{t^{\prime }}\gamma \left( t^{\prime }\right) dt^{\prime
}}\left( C_{\pm }e^{\mp Kt/2}\pm K^{-1}e^{\pm Kt/2}\right) .
\end{equation}
For this case the value of the integration constant $C_{\pm }$ is
\be
C_{\pm }=\frac{2x_{\pm0}}{2v_{\pm 0}+\left(\gamma _0\pm K\right)x_{\pm 0}}\mp K^{-1},
\ee
where $u_{\pm0}=v_{\pm0}/x_{\pm0}$.
Finally, if there is a function $f(t)$ so that the condition
\begin{equation}
\omega ^{2}(t)=\frac{d}{dt}\left[\frac{\gamma (t)\pm \sqrt{f(t)}}{2}\right]+\frac{\gamma
^{2}(t)-f(t)}{4},
\end{equation}
holds for all $t$, then the general solution of Eqs.~(\ref
{ric1}) and (\ref{ric})  are given by
\begin{eqnarray}\label{sol3}
u_{\pm}(t)=e^{\pm \int \sqrt{f(t)}dt}\left[ \int e^{\pm \int \sqrt{f(t)}%
dt}dt+C_{\pm}\right] ^{-1}-
\left[ \frac{\gamma (t)\pm \sqrt{f(t)}}{2}\right] ,
\end{eqnarray}
and
\be\label{sol4}
x_{\pm}(t)=x_{\pm0}\exp \left[
\int_{0}^{t}u_{\pm}\left( t^{\prime }\right) dt^{\prime }\right],
\ee
respectively, where $x_{\pm0}$ is an integration constant.

In order to obtain the values of the arbitrary integration constants for the solutions given by Eqs.~(\ref{sol3}) and (\ref{sol4}) from the initial conditions $x_{\pm}(0)=x_{\pm0}$ and $\dot{x}_{\pm}(0)=v_{\pm0}$, and from the initial values of the functions $f(t)$ and $\gamma (t)$,
we denote $\int \sqrt{f(t)}dt=F(t)$, and $\int e^{\pm F(t)}dt=\pm G(t)$, respectively. The initial values of these functions at $t=0$ are $F(0)=F_0$, and $G(0)=G_0$. The $t=0$ value of $f(t)$ is $f(0)=f_0$. Then the values of the arbitrary integration constants $C_{\pm}$ can be obtained as
\begin{equation}
C_{\pm }=e^{\pm F_{0}}\left( \frac{v_{\pm 0}}{x_{\pm 0}}+\frac{\gamma
_{0}\pm \sqrt{f_{0}}}{2}\right)^{-1} \mp G_{0}.
\end{equation}

\subsection{Solitons}

Second order partial differential equations of the form

\begin{equation}
b\left( x,t\right) \frac{\partial ^{2}\Psi (x,t)}{\partial x^{2}}+a\left( x,t\right) \frac{\partial \Psi \left( x,t\right) }{v\partial t}%
+V\left(
x,t\right) \Psi \left( x,t\right) =0,  \label{sol}
\end{equation}
where $v$ is a constant, and $a$, $b$, $V$ are arbitrary functions of the
coordinates $x$ and $t$, are used for the description of a large variety of
physical models. Of particular importance are the so-called soliton
(solitary wave) solutions \cite{sol,sol1}, in which the dependence of the wave function $%
\Psi \left( x,t\right) $ is assumed to be of the form $\Psi \left(
x,t\right) =\Psi \left( vt-x\right) =\Psi \left( \xi \right) $. Therefore
solitons can be obtained as solutions of the second order  ordinary
differential equation
\begin{equation}
b\left( \xi \right) \Psi ^{\prime \prime }\left( \xi \right) +a\left( \xi
\right) \Psi ^{\prime }\left( \xi \right) +V\left( \xi \right) \Psi \left(
\xi \right) =0,  \label{sol1}
\end{equation}
where a prime denotes the derivative with respect to $\xi $.
By dividing this equation with $b(\xi )\neq 0$, and denoting $\gamma \left(
\xi \right) =a\left( \xi \right) /b\left( \xi \right) $ and $\omega
^{2}\left( \xi \right) =V\left( \xi \right) /b\left( \xi \right) $, Eq. (\ref
{sol1}) takes the form of Eq.~({\ref{ric}), the equation of motion of the damped time dependent harmonic
oscillator,

\begin{equation}
\Psi ^{\prime \prime }\left( \xi \right) +\gamma \left( \xi \right) \Psi
^{\prime }\left( \xi \right) +\omega ^{2}\left( \xi \right) \Psi \left( \xi
\right) =0.
\end{equation}

Therefore all the results of the previous subsection can be applied to the
case of the solitons, leading to the possibility of constructing explicit
exact solitonic solutions for the wave-type equations of mathematical
physics.

\section{Conclusions}\label{sect6}

A new method to generate analytical solutions of the Ricacti equation  was presented. The method is based on the correspondence between the initial Riccati equation and a more general equation containing a solution generating function $f(x)$. If the coefficients of the Riccati equation and the function $f(x)$ satisfy one differential and one integral constraints, the general solution of the Riccati equation can be explicitly obtained. The method was illustrated by obtaining the general solution for a number of specific Riccati type equations, with coefficients satisfying the required integrability condition. Some physical applications of the integrability method were considered, by explicitly obtaining several classes of general solutions for the harmonic damped oscillator with time dependent frequency. The possibility of obtaining soliton type solutions of the second order partial differential equations was also briefly considered.

\acknowledgments
We would like to thank to the anonymous referee for comments and suggestions that helped us to significantly improve our manuscript.


\begin{thebibliography}{99}	

\bibitem{1r}  A. D. Polyanin, V. F. Zaitsev, {\it Handbook of exact solutions for ordinary differential equations}, Boca Raton, Chapman \& Hall/CRC (2003).

\bibitem{2r} M. V. Soare, P. P. Teodorescu and I. Toma, {\it Ordinary differential equations with applications to mechanics}, Dordrecht, Springer (2007).

\bibitem{1} F. Cooper, A. Khare, and U. Sukhatme, {\it Supersymmetry and quantum mechanics}, Phys.
Rep. 251, 267 (1995).

\bibitem{2} M. I. Zelekin, {\it Homogeneous Spaces and Riccati Equation in
Variational Calculus}, Factorial, Moscow (1998).

\bibitem{3} V. B. Matveev and M. A. Salle, {\it Darboux Transformations and
Solitons}, Springer, Berlin  (1991).

\bibitem{4} I. L. Buchbinder, S. D. Odintsov, and I. L. Shapiro, {\it Effective
Action in Quantum Gravity}, IOP Publishing Ltd., Bristol (1992).

\bibitem{5} K. Milton, S. D. Odintsov, and S. Zerbini, {\it Bulk versus brane running couplings}, Phys. Rev. D 65, 065012 (2002).

\bibitem{6} H. C. Rosu, F. Aceves de la Cruz, {\it One-parameter Darboux-transformed quantum actions in Thermodynamics},  Physica Scripta 65, 377 (2002).

\bibitem{7}  M. Nowakowski and H. C. Rosu, {\it Newton¡¦s laws of motion in the form of a Riccati equation}, Phys. Rev. E  65, 047602 (2002).

\bibitem{8} P. Olesen, J. Ferkinghoff-Borg, M. H. Jensen, and Joachim Mathiesen, {\it Diffusion, fragmentation, and coagulation processes: Analytical and numerical results}, Phys. Rev. E 72, 031103 (2005).

\bibitem{9} A. Al Bastami, M. R. Belic, D. Milovic, and N. Z. Petrovic, {\it Analytical chirped solutions to the (3 + 1)-dimensional Gross-Pitaevskii equation for various diffraction and potential functions}, Phys. Rev. E 84, 016606 (2011).

\bibitem{8a} H. C. Rosu, O. Cornejo-Perez, and P. Ojeda-May, {\it Traveling kinks in cubic nonlinear Ginzburg-Landau equations}, Phys. Rev. E 85, 037102 (2012).

\bibitem{Ram} J. F. Carinena and A. Ramos, {\it Integrability of Riccati equation from a group theoretical viewpoint}, Int. J. Mod. Phys. A 14, 1935 (1999).

\bibitem{Stel} V. M.  Strelchenya, {\it A new case of integrability of  the general Riccati equation and
its application to relaxation  problems}, J. Phys. A: Math. Gen. 24, 4965  (1991).

\bibitem{Abram} M. Abramowitz, I. A. Stegun, {\it Handbook of mathematical functions with formulas, graphs, and mathematical tables}, Washington, D.C., National Bureau of Standards (1972).

\bibitem{osc} S. I. Denisov, W. Horsthemke, {\it Anomalous diffusion and stochastic localization of damped quantum particles}, Phys. Lett. A. 282, 367 (2001).

\bibitem{osc1} H. Goldstein, C. Poole, J. Safko, {\it Classical mechanics}, San Francisco, Addison Wesley (2002).

\bibitem{Rosu1} H. C. Rosu and M. A. Reyes, {\it Riccati parameter modes from Newtonian free damping motion by supersymmetry}, Phys. Rev. E 57, 4850 (1998).

\bibitem{Rosu2} H. C. Rosu and P. B. Espinoza, {\it Ermakov-Lewis angles for one-parameter supersymmetric families
of Newtonian free damping modes}, Phys. Rev. E 63, 037603 (2001).


\bibitem{sol} T. Miwa, M. Jimbo, and E. Date, {\it Solitons: differential equations, symmetries and infinite dimensional algebras}, Cambridge, Cambridge University Press (2000).

\bibitem{sol1}  M. Dunajski, {\it Solitons, instantons, and twistors}, 	 Oxford, New York, Oxford University Press (2010).

\end{thebibliography}
\end{document}